\documentclass[10pt]{article}
\usepackage{geometry}
\usepackage[all]{nowidow}
\emergencystretch=\maxdimen
\usepackage{amsmath,amssymb}

\usepackage{xcolor}
\definecolor{graycolor}{gray}{0.9} 
\usepackage{microtype}
\usepackage{setspace} 
\usepackage[utf8]{inputenc}
\usepackage[english]{babel}
\usepackage{times}
\usepackage{array}
\usepackage{soul}
\usepackage{cite}
\usepackage[numbers,sort&compress]{natbib}
\usepackage{natbib}
\setcitestyle{open={[},close={]},citesep={,\!},numbers}

\usepackage{titlesec} 
\titleformat {\section} [block] {\raggedright \fontsize{10}{10}\selectfont\bfseries} {\thesection. \space} {0pt} {}
\titlespacing {\section} {0pt} {12pt} {6pt}
\titleformat {\subsection} [block] {\raggedright \fontsize{10}{10}\selectfont\itshape} {\thesubsection .\space} {0pt} {}
\titlespacing {\subsection} {0pt} {12pt} {6pt}
\titleformat {\subsubsection} [block] {\raggedright \fontsize{10}{10}\selectfont} {\thesubsubsection .\space} {0pt} {}
\titlespacing {\subsubsection} {0pt} {12pt} {6pt}
\titleformat {\paragraph} [block] {\raggedright \fontsize{10}{10}\selectfont} {} {0pt} {}
\titlespacing {\paragraph} {0pt} {12pt} {6pt}

\usepackage{array} \newcommand{\PreserveBackslash}[1]{\let\temp=\\#1\let\\=\temp}
\newcolumntype{C}[1]{>{\PreserveBackslash\centering}m{#1}}
\newcolumntype{R}[1]{>{\PreserveBackslash\raggedleft}m{#1}}
\newcolumntype{L}[1]{>{\PreserveBackslash\raggedright}m{#1}}
\usepackage{lineno}
\usepackage{tabularx}
\usepackage{colortbl}
\usepackage{graphicx}
\usepackage{float}
\usepackage[export]{adjustbox}
\usepackage{caption}
\usepackage{fancyhdr} 
\usepackage{lastpage}
\usepackage{layout}
\usepackage{setspace} 
\usepackage{enumitem}
\usepackage{booktabs}
\usepackage{arydshln}
\usepackage{multirow}
\usepackage{color}
\usepackage{hyperref} 
\hypersetup{
	colorlinks=true,
	linkcolor=blue,
	filecolor=blue,
	urlcolor=black,
	citecolor=cyan,
}

\fancypagestyle{firstpage}{
    \setlength{\headsep}{2.2cm}
    
    \setlength{\footskip}{1.5cm}
    \fancyhf{}
    \lhead{\begin{table}[H]
        \centering
        \begin{tabular}{L{2.5cm}C{10cm}C{3.1cm}R{2cm}}
            \includegraphics[scale=0.035]{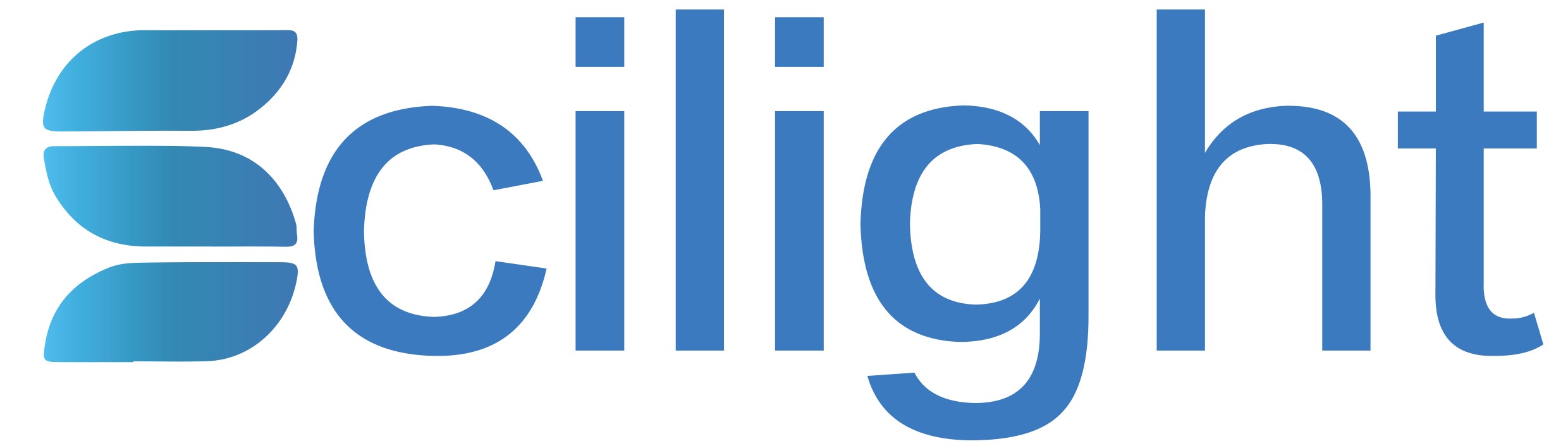} \vspace{-6pt}& \cellcolor{graycolor}\begin{tabular}[c]{@{}c@{}}\textit{Highlights in High-Energy Physics}\\ \href{https://www.sciltp.com/journals/hihep}{https://www.sciltp.com/journals/hihep}\end{tabular} & \includegraphics[scale=0.014]{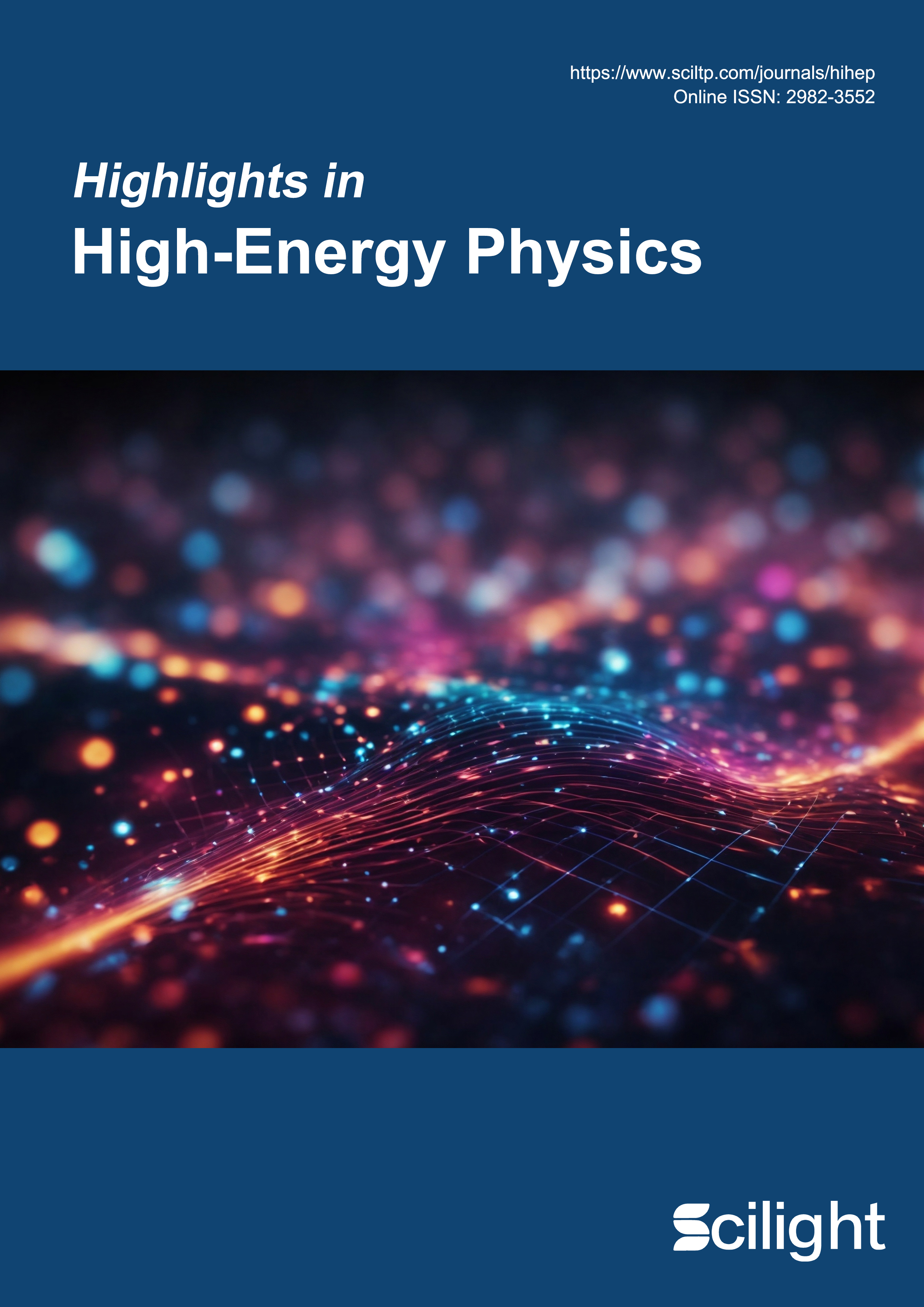} \vspace{-3pt}\\
        \end{tabular}
        \vspace{-22pt}
    \end{table}}
   
    \fancyfoot[C]{
        \vspace{-1.55cm}
        \begin{table}[H]
            \begin{minipage}[c]{0.15\columnwidth}
                \includegraphics[scale=0.5]{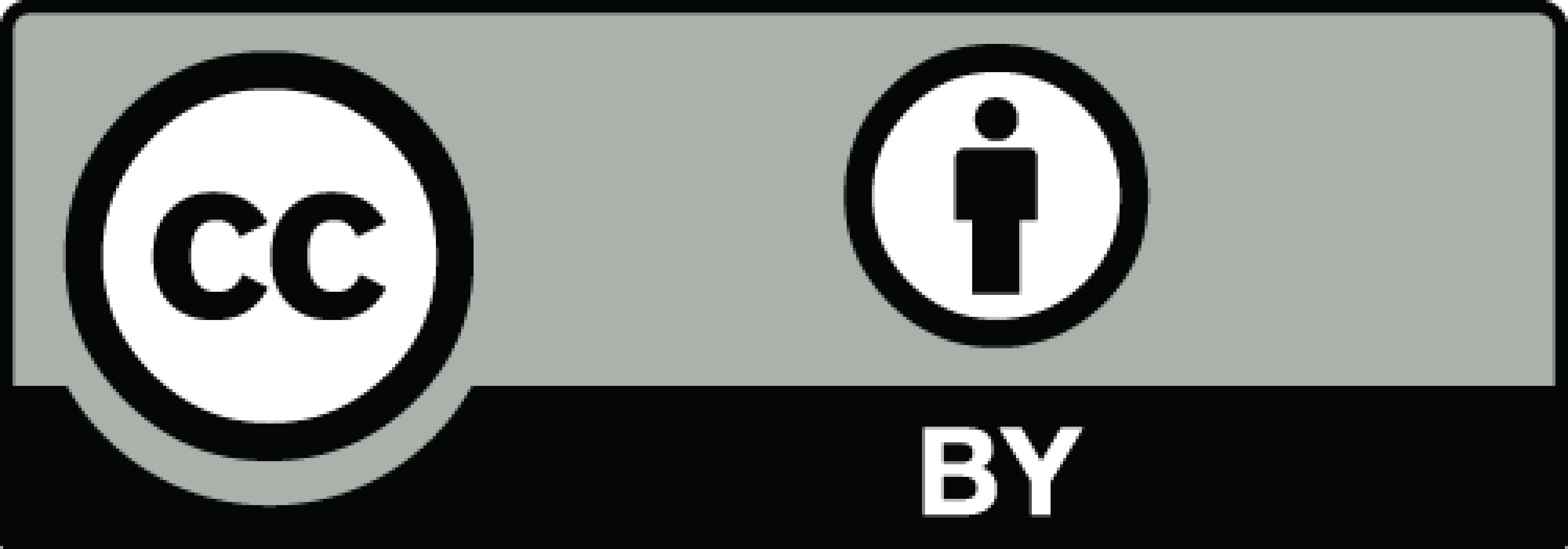} \vspace{1.1pt}
            \end{minipage}
            \hfill
            \begin{minipage}[c]{0.85\columnwidth}
                \scriptsize \textbf{Copyright:} © 2026 by the authors. This is an open access article under the terms and conditions of the Creative Commons Attribution (\mbox{CC BY}) license (\href{https://creativecommons.org/licenses/by/4.0/}{https://creativecommons.org/licenses/by/4.0/}). \\ \textbf{Publisher’s Note:} Scilight stays neutral with regard to jurisdictional claims in published maps and institutional affiliations.
            \end{minipage}
    \end{table}}
    \vspace{-0.55cm}
}

\begin{document}
\newgeometry{left=2.5cm, right=2.5cm, top=1.8cm, bottom=4cm}
	\thispagestyle{firstpage}
	{\noindent \textit{{Article}  
}}
	\vspace{4pt} \\
	{\fontsize{18pt}{10pt}\textbf{Heavy Particle Towers and Nonlocal QFT}  }
	\vspace{16pt} \\
	{\large Stathes Paganis \textsuperscript{1,2}}
	\vspace{6pt}
	 \begin{spacing}{0.9}
		{\noindent \small
			\textsuperscript{1}	\parbox[t]{0.98\linewidth}{\textls[-15]{Department of Physics, National Taiwan University, No 1, Sec 4, Roosevelt Road, Taipei 10617, Taiwan; paganis@phys.ntu.edu.tw}} \\
			\textsuperscript{2}	Leung Center for Cosmology and Particle Astrophysics,
National Taiwan University, Taipei 10617, Taiwan \vspace{6pt}\\
		\footnotesize	\textbf{How To Cite}: Paganis, S. Heavy Particle Towers and Nonlocal QFT. \emph{Highlights in High-Energy Physics} \textbf{2026}, \emph{Volume}(Issue), Page Number. \href{https://doi.org/10.xxxx/xxx}{https://doi.org/10.xxxx/xxx}}\\
	\end{spacing}

\begin{table}[H]
\noindent\rule[0.15\baselineskip]{\textwidth}{0.5pt} 
\begin{tabular}{lp{12cm}}  
 \small 
  \begin{tabular}[t]{@{}l@{}} 
  \footnotesize  Received: day month year \\
  \footnotesize  Revised: day month year \\
   \footnotesize Accepted: day month year \\
  \footnotesize  Published: day month year
  \end{tabular} &
  \textbf{Abstract:} A number of gravitation-motivated theories, as well as theories with new coloured fermions predict heavy particle towers with spectral densities $\rho(m^2)$ growing faster than $e^{m}$, a 
characteristic of nonlocalizable theories. It is shown that if a light scalar, like the Higgs boson, interacts strongly with a heavy scalar particle tower with exponentially rising degeneracy, then the local low-energy theory is equivalent to an effective nonlocal scalar QFT. 
For energies approaching the nonlocality scale $p^2\simeq\Lambda_{NL}^2$, the scalar propagator and scattering amplitudes are modified by nonlocal factors of the form $e^{p^2/\Lambda^2_{NL}}$. The double-Higgs production measurement at the LHC is proposed as a highly sensitive probe of nonlocality at the electroweak scale. \\
\\
  & 
  \textbf{Keywords:} {Higgs boson; BSM; Nonlocal field theories}  
 
\end{tabular}
\noindent\rule[0.15\baselineskip]{\textwidth}{0.5pt} 
\end{table}

	\section{Introduction }\label{intro}
	Towers with a high density of energy levels may display an exponentially increasing degeneracy of states with energy scale $p^2$. Theories associated with such towers may be nonlocalizable~\cite{Galileon2015,Buonin2022}. For instance, since strings are inherently nonlocal structures, models inspired by string field theory are expected to lead to violations of locality. In this case, for energy scales $p^2\sim\mathrm{TeV}^2$ approaching the scale of nonlocality $\Lambda_{NL}^2$ from the IR, the SM interaction vertices would appear effectively smeared. This smearing would lead to modifications of SM scattering amplitudes by form factors that could be measured at the LHC~\cite{BISWAS2015113}. 
The observed value of the muon anomalous moment can also be explained in the framework of nonlocal theories~\cite{Capolupo:2022awe}.
To study such effects, one can use a nonlocal Lagrangian that describes the interaction of the SM particles and the BSM heavy tower, which for $p^2 << \Lambda_{NL}^2$ should recover the SM Lagrangian. 
An interesting connection between particle towers and nonlocality has been presented in~\cite{PhysRevD.104.015028}, where it is argued that asymptotically nonlocal field theories interpolate between Lee-Wick theories with multiple propagator poles and ghost-free nonlocal theories.

\textls[-15]{Our work is an attempt to link nonlocalizable theories with SM phenomenology 
\cite{Paganis:2026ota}. Heavy particle towers associated with nonlocalizable theories, when coupled to SM particles, could lead to exponentially rising scattering amplitudes with the external momentum $p^2$. Nonlocality emerges as $p^2$ approaches the new scale $\Lambda_{NL}^2$.
Instead of {\textit{asymptotically nonlocal}}
, we will still characterize this behaviour as \textit{nonlocal} since it is sourced by a nonlocalizable~theory.}

The nonlocal nature of a theory can be studied using the spectral density of states $\rho(m^2)$ of the Hamiltonian, defined by:
\begin{equation}
    \sum_n \langle 0 |\mathcal{O}|n\rangle\langle n |{\mathcal{O}^\dagger}|0\rangle
    (2\pi)^4 \delta^{(4)}(m^2-m_n^2)= 2\pi\rho_{\mathcal{O}}(m^2),
    \label{rhodefinition1}
\end{equation}
where $|n\rangle$ is a complete set of states with all particles in state $n$ being on shell, and $|\langle 0|\mathcal{O}|n\rangle|^2$ the probability that operator $\mathcal{O}$ produces a state $n$ at rest, with mass $m_n$.
In a local theory the spectral density is polynomial bounded~\cite{Jaffe:1967un}.
Nonlocalizable theories are characterized by
an exponential increase, $\sim e^m$, of the spectral density. The full scattering amplitudes on the other hand, need to be regularized through a subtraction procedure and be finite at the ultraviolet (UV). Violations of locality appear 
in some UV completions such as string theory at the perturbative level 
\cite{deRham:2010kj,deRham:2014zqa,Cheung:2014dqa,Nortier:2021six}, because of non-perturbative effects in the classicalization proposal~\cite{Cachazo:2014xea}, as well as in \textls[18]{Galileon theories~\cite{Hinterbichler:2015pqa}. It has also been conjectured that the UV description of Galileons, and by extension all} 
\restoregeometry

\noindent  theories of Massive gravity, Bi-gravity and Multi-gravity violate the condition of polynomial boundedness, but they respect Unitarity, Analyticity, and Lorentz Invariance~\cite{Galileon2015}.

Nonlocal QFTs (NLQFT)~\cite{Tomboulis:1997gg,Modesto:2011kw,Ghoshal:2017egr,Buoninfante:2018mre,Ghoshal:2020lfd,Nortier:2023dkq,Chattopadhyay:2023nbj,Alexeyev:2017scq,Shimada:2025rpt,Maiezza:2019dht,El-Nabulsi:2013cta,Tomboulis:2015gfa}
are nonlocal theories that employ entire transcendental functions of higher-derivative quadratic terms, $\mathcal{F}(\partial_{\mu}\partial^{\mu})$, to improve the convergence of loop amplitudes, without introducing any ghosts.
A general NLQFT action for a single 
scalar field $\phi(x)$ is given by~\cite{Buoninfante:2018mre}:
\begin{equation} 
\mathcal{A} = \int d^4x\left[-\frac{1}{2}\phi \mathcal{F}(\Box) \left(\Box+m_{\phi}^2\right) \phi - V\left(\phi\right)\right],
\label{NLagrangian} 
\end{equation}   
where $\Box=\partial_{\mu}\partial^{\mu}$, and operator $\mathcal{F}(\Box)$ is an infinite series of derivatives. 

In this work, we will employ $\mathcal{F}(\Box)=e^{\sigma\frac{\Box}{\Lambda_{NL}^2}}$, because in many nonlocalizable models the spectral density increases as $e^{\sigma p^2}$.
Taking $\sigma=1$, or equivalently absorbing this factor to the new physics scale, the Action becomes:
\begin{equation} 
\mathcal{A} = \int d^4x\left[-\frac{1}{2}\phi e^{\frac{\Box}{\Lambda_{NL}^2}} \left(\Box+m_{\phi}^2\right) \phi - V\left(\phi\right)\right].
\label{NLagrangian2} 
\end{equation}
{This}  
 particular non-local scalar action is widely used in the literature, because the entire exponential function $e^{-p^2/\Lambda_{NL}^2}$ has no zeros for any finite complex value of $p^2$, hence its inverse introduces no new poles to the propagator. This ensures that the spectrum is completely free of ghosts, preserving the unitarity of the $S$-matrix while rendering loop diagrams UV-finite in the Euclidean regime. Constant $\Lambda_{NL}$ represents the scale at which non-local effects become important.

Employing a $(+,-,-,-)$ metric, $p^2=-\Box$, the propagator is given by~\cite{Buoninfante:2018mre}:
\begin{equation} 
\Delta\left(p^2\right) =  
\frac{e^{p^2/\Lambda_{NL}^2}}{p^2-m^2_{\phi}+i\epsilon}. 
\label{prop1}       
\end{equation}   
{In} the limit $p^2/\Lambda_{NL}\rightarrow 0$ the local 
Lagrangian and the scalar Feynman propagator are recovered.
The presence of nonlocal operators of the form 
$e^{\Box/\Lambda_{NL}^2}$ in the kinetic terms of the SM, 
can be tested with LHC data~\cite{BISWAS2015113,Su:2021qvm}. 
They enter as dimensionless form factors that modify SM particle scattering amplitudes:
\begin{equation}
    \frac{\sigma_{Meas}}
    {\sigma_{SM}}=
    e^{\frac{p^2}{\Lambda_{NL}^2}}.
    \label{NLxsection}
\end{equation}
{At} present, measurements at the LHC show absence of such anomalous cross sections and limit nonlocality to scales beyond 
a few TeV~\cite{CMS:2018dqv,CMS:2022tqn,Su:2021qvm}.

The standard choice of exponential, $e^{\sigma p^2}$, operators in the literature,~\cite{Buoninfante:2018mre}, lacks physics motivation. 
As already mentioned, in this work we motivate the presence of nonlocal operators of exponential form, through coupling of SM particles to heavy particle towers appearing in a number of nonlocalizable theories. We will not attempt to single out a particular theory, instead we will classify a set of theories according to their spectral densities $\rho(m^2)$. To do so, we will employ a general dispersive approach as in~\cite{Banks:2020gpu}.
When SM particles like the Higgs scalar $h$ couple to a heavy particle tower with exponentially rising degeneracy, 
interactions are effectively nonlocal with the nonlocal effects emerging as $p^2$ approaches the mass of the lightest state from the IR. 
The emergence of nonlocality is due to the exponential increase of degeneracy of real (on-shell) propagating modes with the external momentum $p$.
The nonlocalizability of these models leads naturally to $e^{p^{2\alpha}}$ operators with $1/2 \leq \alpha \leq 1$.

To demonstrate the connection between heavy particle towers 
and nonlocality, we show that  
in the presence of a tower of $n$ interacting scalars with spectral 
density $\rho(m^2)\sim e^{m^2}$ and large number of exchanged modes $n$,
the two-point correlation function behaves asymptotically in the IR as a nonlocal scalar propagator of scalar mass $m_{h}$:
\begin{equation} 
\Delta\left(p^2\right) = \int_{0}^{\infty} \frac{\rho\left(m^2\right)dm^2}{p^2-m^2+i\epsilon}\                   
\xrightarrow{n\rightarrow \infty}  
\frac{e^{p^2/\Lambda_{NL}^2}}{p^2-m_{h}^2+i\epsilon} 
\label{KL1}       
\end{equation} 
{As} a result, scalar amplitudes with ingoing and outgoing Higgs boson lines are modified by exponential form factors, leading to a rich phenomenology at the LHC and beyond.

The paper is organized as follows: in Section \ref{sec:level2} we discuss the classification of theories as local or nonlocalizable. In Section \ref{Towers} we summarize a recipe for calculating two-point correlation functions from the spectral density.
This is subsequently used to make the connection between heavy particle towers and NLQFT.
Finally, in Section \ref{pheno} we propose potential signals of nonlocality at the LHC.

\section{\label{sec:level2}Spectral Density and Nonlocalizable Theories}
          
The two-point correlation function is a special form of a Green's function  that contains all information on the spectrum of a local quantum field theory~\cite{Dudal:2010wn}. These two-point functions include all elementary propagators of the theory, as well as the two-point functions of any local composite operators, which correspond to propagating bound states. 
In the present work we shall only be interested in two-point functions which can be considered as functions of the external momentum $p^2$, analytically continued to the complex plane with a branch cut on the real axis starting at $p^2 = m_1^2=m_{\phi_1}^2$, which is identified as the threshold for multi-particle production associated with the lowest mass state $\phi_1$. Such a two-point correlation function can be cast in the form of a dispersion relation, that is, an integral representation written in terms of the function’s discontinuities in the complex plane.
Non-perturbative two-point correlation functions have been studied intensively during the early days of QCD~\cite{Shifman:1978bx}, and have been used to make predictions of multi-quark heavy hadronic state spectra~\cite{Paganis:1997rf,Paganis:1999ux}.
 
Following~\cite{Banks:2020gpu}, we identify the new BSM physics associated with a heavy tower of scalar states in the interaction Hamiltonian that extends the SM. We can allow SM fields  like the Higgs boson $h$, to couple to this new HT sector via the Hamiltonian: $\mathcal{H}_I = \mathcal{H}_{SM}+g\mathcal{O}_{SM}\mathcal{O}_{HT}$, where $g$ the coupling of $h$ to the BSM heavy scalars $\phi_i$.
As an example, 
\begin{equation}
    \mathcal{H}_I =\mathcal{H}_{SM} + g\mathcal{O}_{SM}\mathcal{O}_{HT}=
    \mathcal{H}_{SM} + \sum_{n}\frac{g_n}{\Lambda^{n-2}}h^2\prod_{i=1}^{n} \phi_i.
    \label{Lint}
\end{equation}
{Note} that in general, the BSM heavy tower operator $\mathcal{O}_{HT}=\mathcal{O}_{BSM}$ may also include SM particle exchanges:
\begin{equation}
    \mathcal{O}_{SM}\mathcal{O}_{HT}= 
    \frac{1}{\Lambda^{n-2}}h^2\prod_{i=1}^{n-k}\phi_i 
    \prod_{j=1}^{k}h_j.
    \label{Lint2}
\end{equation}
{One} can then write down $n$-point scattering amplitudes with the 
Higgs boson as external lines. 
The two-point correlation function in momentum space 
with external momentum $p^2$, is given by:
\begin{equation}
\Delta(p^2) =\langle \mathcal{O}_{HT}(p)\mathcal{O}_{HT}(-p) \rangle. 
\end{equation}     
{In} our example interaction of Equation~\eqref{Lint}, the correlation function describes the exchange of $n$ BSM bosons $\phi_1, \phi_2, ..., \phi_n$, i.e., a $n-1$ loop process. We will later use this example to estimate modifications in the SM double-Higgs production.

Since the early days of QFT, it was realized that the two point-correlation
function admits a general integral representation in terms of a linear 
combination of the Feynman propagator weighted by a spectral density,
$\rho(m^2)$, the K\"{a}ll\`{e}n-Lehmann (KL) spectral representation~\cite{Kallen:1949bza,Lehmann:1954xi}:   
\begin{eqnarray}
\Delta\left(p^2\right)   
&=& \int_{0}^{\infty} \frac{\rho\left(m^2\right)}{p^2-m^2+i\epsilon} 
dm^2 \nonumber \\
&=& \int_{0}^{\infty} \rho(m^2) \Delta_F(p^2,m^2)dm^2. 
\label{KL2}        
\end{eqnarray}     
{The} KL representation provides the most general way to express the 
propagation of a scalar $h$ in the presence of interactions 
described by a Hamiltonian with a mass spectrum of eigenstates 
defined by the spectral density $\rho_{\mathcal{O}}(m^2)$ of Equation~\eqref{rhodefinition1}, which is a real and positive function of $m^2$.
The field $h$ that we will identify with the SM Higgs boson, appears with a mass gap $m_1-m_h$. It should be noted that the tower states may be resonances with poles away from the real $p^2$ axis leading to a smooth change in the spectral density below threshold. This is the most general case where the mass $m$ in the propagator in Equation~\eqref{KL2} has an imaginary part.

%
The growth of spectral densities of fields with $m^2$ provides information on the localizability of the theory. A particular theory (of scalars in our case) is called localizable if $\rho$ is bounded at the UV by a finite-order polynomial. We typically characterize a theory as {\textit{strictly localizable}, \textit{quasi-local} or \textit{nonlocalizable}} by expressing the high-$m^2$ behaviour of the spectral density as~\cite{Galileon2015,Tomboulis:2015gfa,Buonin2022}:  
\begin{equation}
\rho \sim e^{m^{2\alpha}}\times \textrm{subdominant terms}. 
\label{rhoNLclassif}
\end{equation}    
{The} localizability of a theory is then characterized as follows:
\begin{itemize}[align=parleft,leftmargin=*,labelsep=6.5mm]
    \item $\alpha < 1/2$: the theory is strictly localizable,\vspace{-6pt}
    \item $\alpha = 1/2$: the theory is quasi-local,\vspace{-6pt}
    \item $\alpha > 1/2$: the theory is nonlocalizable.
\end{itemize}

In a gravitational theory the spectral densities of generic  
operators grow faster than linear exponentials, i.e. operator-valued fields in quantum gravity are fundamentally nonlocalizable~\cite{Galileon2015}.
To give some examples, in a gravitational theory we expect the density of states at high energies to be determined by the density of states for black holes which scales as    
\begin{equation}   
\rho(m^2) \sim e^{S_{\textrm{BH}}} = e^{c(m/M_{Pl})^{\frac{d-2}{d-3}}}, 
\end{equation}     
where $S_{BH}$ the Beckenstein-Hawking entropy, and $d$ the dimensionality of
spacetime. So, for $d\geq 4$ we get $\alpha \geq 1$, hence the theory is nonlocalizable. 

Little String Theories,~\cite{Berkooz:1997cq,Seiberg:1997zk}, can also be shown to have the exponential growth characteristic of the Hagedorn density of states~\cite{Galileon2015,Aharony:1998tt,Peet:1998wn,Minwalla:1999xi}.
In particular the KL spectral density for the two-point function is 
expected to grow as: 
\begin{equation}
\rho(m^2) \sim e^{\frac{cM_{\textrm{string}}m} {\sqrt{N}}},
\end{equation}    
where $M_{\textrm{string}}$ is the string scale and $N$ the number of five-branes. Thus, Little String Theories are quasi-local.

\section{\label{Towers}Towers of Heavy Scalars and Nonlocality}
As we saw in the previous section, the growth of the spectral density of a theory with $m$ could render the theory nonlocalizable. Our focus in this work is on the phenomenological implications of the emergence of nonlocality, in terms of modifications of low energy effective Lagrangians like the SM.

\subsection{Correlation Functions from Spectral densities}
The relation between the spectral density and the two-point correlation function in Equation~\eqref{KL2}, is provided by application of the Cauchy theorem~\cite{Zwicky:2016lka}:  
\begin{equation}   
\Delta\left( p^2 \right)=-\frac{1}{2\pi i} 
\int_{m^2_{h}}^{\infty}
\frac{\textrm{Disc}\left[ \Delta\left(m^2\right) \right]}{p^2-m^2+i\epsilon} dm^2,                          
\label{eq:disc}
\end{equation}
where the integrand contains the discontinuity of the function $\Delta(p^2)$ across the cut on the real axis for $m^2>m^2_{h}$. The discontinuity is given by: 
\begin{equation}  
\textrm{Disc}\left[\Delta\left(m^2\right) \right] =  
2i\textrm{Im}\left[\Delta\left(m^2\right) \right].  
\label{eq:disc2}
\end{equation}     
{Replacing} in Equation~\eqref{eq:disc}, we obtain:  
\begin{equation}      
\Delta\left(p^2\right)=-\frac{1}{\pi}  
\int_{0}^{\infty}         
\frac{\textrm{Im}\left[\Delta\left(m^2\right) \right]}{p^2-m^2+i\epsilon}dm^2.
\end{equation}     
{By} comparison with Equation~\eqref{KL2}, the scalar spectral density is given by:
\begin{equation} 
\rho\left( m^2 \right) = - \frac{1}{\pi}\textrm{Im}\left[\Delta\left(m^2\right) \right].      
\label{eq:rho1} 
\end{equation}     
{This} is the familiar result that the spectral density is proportional to the imaginary part of the scattering amplitude, and is what we should expect for scattering involving exchanges of real (on-shell) particles.

The forward scattering amplitude 
$\mathcal{M}$$(A\rightarrow X_n \rightarrow A)$ of an initial state 
of SM particles $A$, where $X_n$ denotes the exchange of $n$
real scalars, is given by the optical theorem as:
\begin{eqnarray}   
&&2\mathrm{Im}[\mathcal{M}\left(A\rightarrow A \right)] =\nonumber\\
&&            
\sum_{X_n}       
\int d\mathrm{\Pi}_{X_n} 
|\mathcal{M}\left(A\rightarrow X_n \right)|^2 
\left(2\pi\right)^4 \delta^4\left( p_A - p_{X_n} \right),\nonumber 
\end{eqnarray}   
where              
$d\mathrm{\Pi}_{X_n}=\prod_{i=1}^{n}\frac{d^3p}{(2\pi)^3}\frac{1}{2E_i}$
is the phase space of the exchanged real states. 
In our notation, the relation between the invariant Feynmann amplitude $\mathcal{M}(i\rightarrow f)$  
and the two-point correlation function is:
\begin{equation}                  
\mathcal{M} \left(A \rightarrow A \right) = -\Delta( p^2 ).
\end{equation}     
{Replacing} the amplitude in Equation~\eqref{eq:rho1} 
and using the optical theorem, we can obtain the following expression for the spectral density for $n$-scalar exchange:  
\begin{eqnarray}
&&\rho_n(p_A^2) = \frac{1}{\pi} \textrm{Im}\left[\cal{M} \right] = \nonumber\\
&&\left(2\pi\right)^3 
\int d\mathrm{\Pi}_{X_n} 
|\mathcal{M}\left(A\rightarrow X_n \right)|^2 
\delta^4\left( p_A - p_{X_n} \right).
\label{Opt2}
\end{eqnarray} 
{For} the tree-level amplitude $\mathcal{M}\left(A\rightarrow X_n \right)$, Equation~\eqref{Opt2} provides a means of extracting the spectral density $\rho_{\mathcal{O}}(p^2)$ of a particular operator $\mathcal{O}$~\cite{Banks:2020gpu}. Since the exchange of $n$-scalars corresponds to $n-1$ loops, for $n=2$ Equation~\eqref{Opt2} would relate the cross section of the cut diagram $A\rightarrow \phi_i\phi_j$ with the spectral density of the operator 
$\mathcal{O}_{HT}=\phi_i\phi_j$. Examples of such operators are shown in Equations~\eqref{Lint} and \eqref{Lint2}. The spectral density can be then used to obtain two-point correlation functions using the KL representation, Equation~\eqref{KL2}.

\subsection{Towers and Nonlocalizability}
As an elementary example of the dispersive approach presented above, we calculate the spectral density for the exchange of $n$ identical tower scalars $\phi$ of mass $m_{\phi}$ in the Higgs-Higgs fusion process:
\begin{equation}
    hh\rightarrow \phi^n \rightarrow hh.
\end{equation}
{This} is an $n-1$ loop diagram. The leading-order SM diagram of this process, is the Higgs-boson tree-level exchange.
We can allow $n$ to vary by introducing in the Lagrangian dimension $n+2$ operators of the form:
\begin{equation}
    \mathcal{O}_{SM}\mathcal{O}_{HT}=\frac{1}{\Lambda^{n-2}}hh \phi^n,
\label{Oper1}
\end{equation}
where $g_n\sim 1$, and we have explicitly identified the exchanged scalars $\phi$ as the BSM sector. As shown in Equation~\eqref{Lint2}, we can allow for $h$ itself to be exchanged by introducing operators 
$\mathcal{O}_{SM}\mathcal{O}_{HT}=\frac{1}{\Lambda^{n-2}}hh \phi^{n-k}h^k$. Although this choice will affect the phenomenology, working for simplicity with the operator in Equation~\eqref{Oper1}, will not change our main conclusions.
The branching amplitude 
$\mathcal{M}\left(A\rightarrow X_n \right)$ in Equation~\eqref{Opt2}, is simply given by $n!$~\cite{Banks:2020gpu}:
\begin{equation}
\mathcal{M}\left(hh\rightarrow \phi_n\right) = n!    
\end{equation}     
{Thus}, in this case the spectral density can be written as: 
\begin{align}   
\rho_n\left( m^2 \right) &= \frac{(n!)^2 }{2\pi}\left[\frac{1}{n!}  
\int d\mathrm{\Pi}_{X_n} 
\left(2\pi\right)^4 \delta^4\left( m - \sum_i^n p_{i}\right)\right]\nonumber\\
&=\frac{n!}{2\pi}I_n,   
\label{eq:rho4}    
\end{align}        
where the $1/n!$ factor is applied in the case of exchange of identical bosons, and the phase-space integral in Equation~\eqref{eq:rho4} is denoted by $I_n$.

We can use Equation~\eqref{eq:rho4} to estimate the spectral density of 
$\mathcal{O}_{HT}$ in Equation~\eqref{Oper1} for a particle tower with energy levels $E^2_k=m^2_k$. A simple expression for the energy levels may be given by:
\begin{equation}
    E_k = m_{1} + (k-1)^\beta \Delta m, ~~~~~k=1,2,...,
\label{elevels}
\end{equation}
where $\beta$ a positive real and $\Delta m$ a level mass-gap parameter. For $\beta=1$, and external momentum $p^2$, the largest number of identical scalars that can be exchanged is
\begin{equation}
    n=\sqrt{p^2-m_1^2}/m_1 
    \xrightarrow{p^2 >> m_1^2} p/m_1.
\end{equation}
{This} is the case where all scalars exchanged are those of the lowest mass, $m_1$ (or $m_h$ in the general case).
Kinematically, all possible combinations of different scalar exchanges can occur, as long as the sum of their masses does not exceed the external momentum $p$. The presence of $\Theta(p-\sum_i^n m_i)$ is implied in Equation~\eqref{eq:rho4}. 
The largest contributions to $\rho$ for a particular $p^2=m^2$ come from 
the degeneracy factor $n!$, which is due to the exchange of $n$ identical scalars. The full spectral density can then
be approximated as: 
\vspace{-6pt}
\begin{equation}  
\rho(m^2) \simeq \sum_{k=1}^n \rho_k(m^2) = \sum_{k=1}^n \left( \frac{k!}{2\pi}I_k \right).
\end{equation}   
{In} this approximation we have ignored as subleading mixed particle degeneracies, as for instance, $n_1$ particles of type $\phi_1$ and $n_2$ particles of type $\phi_2$, and so on.
For large $n$ we can employ the Stirling formula that approximates 
$\sum (n!)$, to write the inequality: 
\begin{equation}
\rho(m^2) > \sum_{k=1}^n (k!) \simeq \sqrt{2\pi n}\left(\frac{n}{e} \right)^n                 
\end{equation}   
or in an exponential form: 
\begin{equation}   
\rho(m^2) > e^{nln(n)} ,
\label{eq:rholast}
\end{equation}     
where we have also omitted the contribution of $I_n$. 
We observe that for
large $n$ the spectral density increases faster than $e^n$, and 
given that subleading contributions are relatively small, $\rho$ 
cannot grow faster than $e^{n^2}\simeq e^{p^2/\Lambda^2_{NL}}$. The theory is clearly 
nonlocalizable since in this case the $\alpha$ parameter of  
Equation~\eqref{rhoNLclassif} lies in the range $1/2\leq \alpha <1$. 
In the expression for the energy levels, Equation~\eqref{elevels}, larger values of the parameter $\beta > 1$ give larger degeneracy and thus faster growing spectral density, also leading to nonlocalizable theories.

\subsection{From Spectral Densities to NLQFT}
For  simple NLQFT Lagrangians as the ones discussed in the 
introduction
\begin{equation}   
\mathcal{L} = -h(x) e^{\frac{\Box}  
{\Lambda_{NL}^2}} \left(\Box +m_h^2 \right) h(x) - V(h),
\label{NLLagrangian}
\end{equation}     
the free propagator is multiplied by a dimensionless
exponential form factor $e^{p^2/\Lambda_{NL}^2}$:  
\begin{equation}   
\Delta_{NL}\left(p^2\right) = 
e^{\frac{p^2}{\Lambda_{NL}^2}}
\frac{1}{p^2-m_h^2+i\epsilon}.
\label{NLrho}   
\end{equation}     
{In} the case of the Higgs-Higgs fusion process mediated by a single-Higgs exchange, and at higher $p^2$
by the exchange of a heavy scalar tower $(\phi_1, \phi_2, ..., \phi_n)$ of spectral density $\rho_{\phi^n}(m^2)$, the two-point function is: 
\begin{eqnarray} 
&&                 
\Delta\left(p^2\right) = \int_{0}^{\infty} \frac{\rho_{\phi^n}\left(m^2\right)dm^2}{p^2-m^2+\
i\epsilon}      
\nonumber \\
&&                 
\Delta\left(p^2\right)\xrightarrow{n\rightarrow \infty}
e^{p^2/\Lambda_{NL}^2}\int_{0}^{\infty} \frac{f\left(m^2\right)dm^2}{p^2-m^2+i\epsilon},      
\label{fproof1}   
\end{eqnarray}    
where $f(m^2)$ is the mass dependent part of the spectral density at
large $n$, which is a sum of delta functions. 
At the IR, $f(m^2)$ is given by the limit of the two-point 
function, which is just the Feynman propagator:
\begin{eqnarray} 
&&\Delta\left(p^2\right) = e^{p^2/\Lambda_{NL}^2}\int_{0}^{\infty} \frac{f\left(m^2\
\right)dm^2}{p^2-m^2+i\epsilon} 
\nonumber \\      
&& 
\Delta\left(p^2\right) \xrightarrow{p^2<<\Lambda_{NL}^2}\nonumber \frac{1}{p^2-m_h^2+i\epsilon}.  
\label{fproof2} 
\end{eqnarray} 
{This} gives $f(m^2)=\delta(m^2-m_{h}^2)$, leading to a propagator 
modified by an exponential form factor:
\begin{equation}
\Delta\left(p^2\right) = e^{\frac{p^2}{\Lambda_{NL}^2}}
\frac{1}{p^2-m_h^2+i\epsilon}.
\label{fproof3}
\end{equation} 
\textls[-15]{{The} resulting effective Higgs propagator looks identical to the one of a scalar NLQFT. In Equation~\eqref{fproof3}, $m_h$ is the physical mass, and the residue is 1, i.e., the Higgs field is renormalized. 
If we switch off the interaction of the Higgs with the tower, the local Higgs propagator is recovered, meaning that the NL propagator in Equation~\eqref{fproof3} is not the propagator of the free theory.
It should be stressed that when SM multi-Higgs exchanges are not considered, Equation~\eqref{fproof3} is not expected to be valid close to the threshold ($p^2\simeq m_1^2$),  because the large-$n$ condition does not hold.}
\subsection{Nonlocal K\"{a}ll\`{e}n-Lehmann Representation}
In the case of a nonlocalizable theory, 
the KL representation can be written using the spectral density $\rho(m^2)$ in an integral representation with non-local propagators:
\begin{eqnarray}
\Delta\left(p^2\right)   
&=& \int_{0}^{\infty} \frac{\mathcal{F}(p^2-m^2)\rho\left(m^2\right)}{p^2-m^2+i\epsilon} 
dm^2\\
&=&\int_{0}^{\infty} \frac{e^{\frac{p^2-m^2}{\Lambda_{NL}^2}}\rho\left(m^2\right)}{p^2-m^2+i\epsilon} 
dm^2.
\end{eqnarray}
{The} IR limit of this definition of the KL, approaches Equation~\eqref{fproof3}.

\subsection{Discussion}
The importance of this result can be understood by inspecting the structure of the interaction terms in the Lagrangian (Equation~\eqref{Oper1}), where the 
pure SM piece carries the external momentum $p^2$. The $e^{p^2/\Lambda^2_{NL}}$ modification represents the effect 
of the $\mathcal{O}_{BSM}$ at the IR: SM scattering amplitudes with ingoing and outgoing Higgs scalars are modified by the same form factor~\cite{BISWAS2015113}.
The resulting exponential increase should hold up to some scale $p^2 \sim\Lambda^2=\Lambda^2_{reg} > m_1^2$ beyond which the amplitudes should be regularized. Scales $p^2 < \Lambda^2_{reg}$ are relevant to SM phenomenology since, for example, Higgs-Higgs scattering amplitudes should receive significant corrections.


Interestingly, in the case of certain Lee-Wick theories~\cite{PhysRevD.104.015028}, a similar asymptotic nonlocality behaviour is observed. In our case, we argue that a whole class of nonlocalizable theories with heavy towers of states with spectral densities growing as $e^m$ or faster, should behave like an infinite derivative nonlocal QFT at low energies. As pointed out in~\cite{Carone:2023cnp}, experimentally, an exponential rise of scattering amplitudes should be observed as $\Lambda_{NL}$ is approached from below. The amplitude is expected to stop rising and be regularized at higher energies. This characteristic anomalous rise of the cross section with respect to expectation at lower energies is the smoking gun of the onset of such BSM theories.

This connection between NLQFT, where the non\-locality-inducing entire functions $\mathcal{F}(\Box)$ are arbitrarily inserted in the Lagrangian, and a class of models which naturally lead to $\mathcal{F}(\Box)\sim e^{\Box}$ form factors, is of particular interest. It should be stressed that one would still need a physical mechanism to explain how to generate the desired hierarchies so that the scale of nonlocality is not beyond experimental reach. 

\section{\label{pheno}Collider Phenomenology}

LHC searches for heavy resonances and continuum excesses 
in diboson final states or states with boosted objects in the final state~\cite{ATLAS:2023hbp,CMS:2022tqn,CMS:2018dqv,Li:2019pag,Hoffmann:2014aha}, are just a few examples of attempts to probe models that predict towers of states
\cite{Arkani-Hamed:1998jmv,Randall:1999ee,Giudice:2016yja,Branco:2011iw}.
The main observable of nonlocality is the exponential  
modification of the measured cross section with respect to the expectation from SM, as the external scale approaches the NL scale $p^2\rightarrow \Lambda^2_{NL}$~\cite{BISWAS2015113,Su:2021qvm}:
\begin{equation} 
\frac{\sigma^{Meas}(X\rightarrow Y)}{\sigma^{SM}(X\rightarrow Y)} = e^{\frac{p^2}{\Lambda^2_{NL}}}.
\label{Pheno1}
\end{equation} 

An application of these ideas at the LHC relates to measurements of the $W^{\pm}, Z$ and Higgs-boson final states: $Vh, hh, VV$, 
where $V=W^{\pm}, Z$ represents the weak gauge bosons. When these final states are produced via the Higgs Vector-Boson Fusion (VBF) or Vector-Boson Scattering (VBS) production modes ($VV\rightarrow Vh, hh, VV$), the corresponding scattering amplitudes can receive contributions from nonlocal effects leading to deviations from the SM described by Equation~\eqref{Pheno1}.

\begin{figure}[H]      
\centering 
\includegraphics[width=0.98\textwidth]{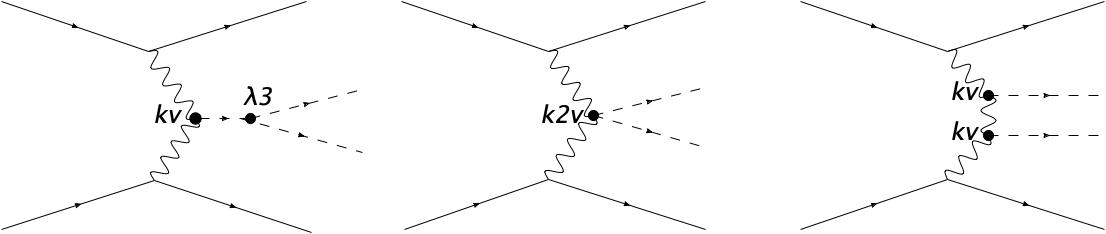}
\caption{VBF double-Higgs production tree-level graphs: $s$-channel Higgs (\textbf{left}), quartic $VVhh$ channel (\textbf{center}), and $t$-channel $V$ exchange (\textbf{right}).} 
\label{fig:VBFtoDiHiggs2}
\vspace{-18pt}
\end{figure}

The leading-order (LO) double-Higgs VBF production Feynman diagrams are shown in Figure~\ref{fig:VBFtoDiHiggs2}. For large scales $p^2>>M_W^2$ and in the absence of any new BSM physics scale $\Lambda_{BSM}$, $p^2<<\Lambda_{BSM}^2$, the longitudinal vector boson scattering dominates the amplitude~\cite{Contino:2010rs}. The LO amplitude in the limit $m_W^2<<p^2$ is given by~\cite{Bishara:2016kjn}:
\begin{equation}
    \mathcal{A}(V_LV_L\rightarrow hh)\simeq \frac{s}{v^2}\left(k_{2V} - k_{V}^{2} \right),
\label{SMVVhhLO}
\end{equation}
where $s=p^2$ is the square of the center of mass energy. 
The measured couplings $k_{2V}$, $k_{V}$ and $\lambda_3$ shown in Figure~\ref{fig:VBFtoDiHiggs2}, are all normalized to their SM values: 
\begin{equation}
k_{2V}=\frac{g_{2V}^{Meas}}{g_{2V}^{SM}},
~k_{V}=\frac{g_{V}^{Meas}}{g_{V}^{SM}},
~\lambda_3=\frac{\lambda_{3}^{Meas}}{\lambda_{3}^{SM}}.
\label{vbfhhprod}
\end{equation}
{In} Equation~\eqref{SMVVhhLO}, only the quartic $VVhh$ diagram and the $t$-channel $V$ exchange are considered. The $s$-channel Higgs exchange is subleading at high $s$ and is neglected. Although the $V_LV_L\rightarrow hh$ scattering is rising with $s$, for SM couplings $k_{2V}=k_V=1$ the amplitude becomes small due to the destructive interference between the quartic and the $t$-channel graphs.

\begin{figure}[H]      
\centering 
\includegraphics[width=0.4\textwidth]{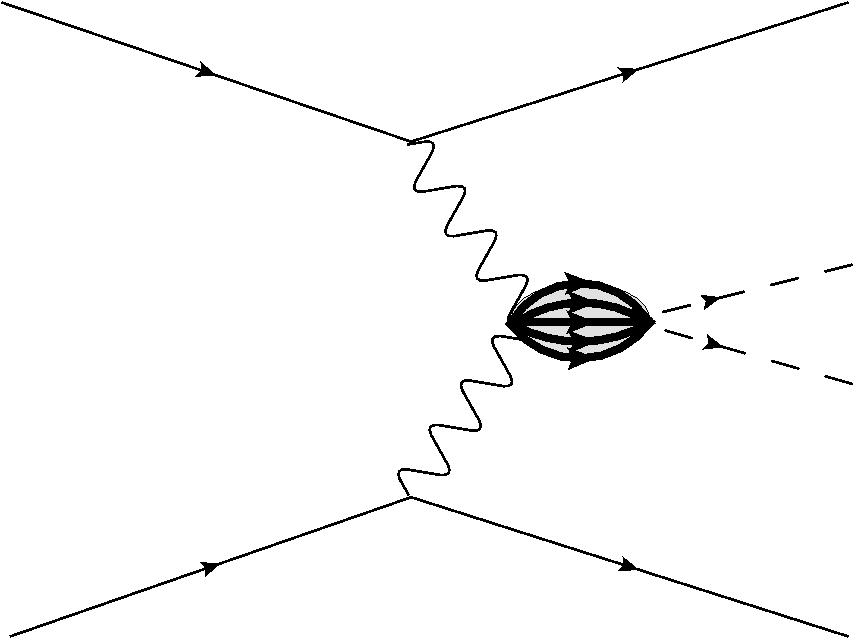}
\caption{Non-local quartic $VVhh$ interaction.} 
\label{fig:VBFtoDiHiggsNL}
\vspace{-12pt}
\end{figure}

Due to this destructive interference, the double-Higgs production through the VBF process at the LHC is extremely hard to measure.
In the case when $s$ is approaching a BSM physics scale $\Lambda_{BSM}^2$,
the scattering amplitude is modified and the cross section receives corrections leading to enhanced double-Higgs rates~\cite{Bishara:2016kjn}.
If the Higgs boson $h$ couples to a heavy scalar tower with nonlocality scale $\Lambda_{NL}$, the quartic coupling will be modified by an exponential form (Figure~\ref{fig:VBFtoDiHiggsNL}). 
The amplitude now becomes, up to $\mathcal{O}(m_W^2/s)$ corrections:
\begin{equation}
    \mathcal{A}(V_LV_L\rightarrow hh)\simeq \frac{s}{v^2}\left(e^{\frac{s}{\Lambda_{NL}^2}} - 1 \right).
\label{NLVVhhLO}
\end{equation}
{Since} the VBF process can be experimentally tagged by requiring two forward-backward quark jets with a rapidity gap, these events provide very clean probes of new physics at the LHC. 
The double-Higgs signal events should display a dijet rapidity gap that increases exponentially with $s$.
The potential of observation of anomalous VBF double-Higgs production in colliders has been studied in~\cite{Bishara:2016kjn} for the case of composite Higgs models. In addition to gravity-inspired models, composite Higgs models predict towers of states. As a result, exponential form-factor modification of the quartic coupling $k_{2V}$ is also expected for a composite Higgs. The details of the increase of the double-Higgs cross section depend on the spectral density of states associated with the model. 

At the LHC and beyond, early signs of emergent nonlocality will appear as growth in the cross section not so different from what is expected from the presence of a resonance with a mass beyond the experimental kinematic reach~\cite{Carone:2023cnp}. The actual form of this growth will depend on the spectral density, so one could imagine a detailed data analysis were spectral density parameterizations are fitted and constrained. The double-Higgs, $Vh$, and $VV$ final states may serve as powerful observables for testing nonlocality~\cite{Su:2021qvm,Chekanov:2021huv}. Since at high energies the $VV$ scattering amplitudes are dominated by the longitudinal degrees of freedom of the weak gauge bosons, we can group all VBF and VBS channels in a single process that needs to be measured: Higgs-Higgs field scattering.

\section{Summary and Conclusions} 
A wide class of theories ranging from gravitation-inspired models to new theories of colour, predict towers of heavy states with spectral densities rising as $e^{m}$ or faster. 
In this work, the connection between such nonlocalizable theories 
and low energy SM phenomenology was addressed.
For a SM Higgs coupled to the heavy tower,
we used the K\"{a}ll\`{e}n-Lehmann spectral representation of the Higgs propagator, to obtain an effective Lagrangian with a nonlocal Higgs-boson propagator of the form:
$e^{p^2/\Lambda^2_{NL}}\frac{1}{p^2-m^2_h+i\epsilon}$, where $m_h$ is the physical Higgs mass.
The resulting effective NL Lagrangian leads to modifications of scattering amplitudes by an exponential form factor $e^{p^2/\Lambda^2_{NL}}$. 
Nonlocal effects emerge at a scale $\Lambda_{NL}$ that is in general expected to be close to the mass threshold of the tower $m_1$, albeit models with $\Lambda_{NL}<<m_1$ have also been proposed~\cite{PhysRevD.104.015028,Boos:2021lsj}.
At the IR limit, $p^2 << \Lambda^2_{NL}$, the scattering amplitudes approach the SM ones. At the UV, the exponential growth of the scattering amplitudes is expected to be regularized at a scale $\Lambda_{reg}$.

It is rather interesting that nonlocalizable theories can induce nonlocal effects at TeV scales. As we have shown, all information about these effects is contained in the spectral density $\rho(m^2)$. We argued that LHC measurements of certain scattering amplitudes can provide constraints on the spectral density. As an example, we proposed and briefly discussed the Higgs-Higgs (field) scattering at very high energies at the LHC. If the nonlocality scale, $\Lambda_{NL}$, is of order $\mathcal{O}(1-10~TeV)$, the Vector Boson Fusion double-Higgs production rate should be significantly increased. Increases are also expected in the VBF single-Higgs production, as well as in the Vector Boson Scattering process cross section. Observation of exponential rise of the double-Higgs or di-Boson cross sections with respect to the SM expectation, will be evidence of nonlocality.
	

 

		\section*{Funding}
{This work was supported by the Taiwanese NSTC grant 112-2112-M-002-026-MY3. }  




		\section*{Data Availability Statement}
 {Not applicable}
		\section*{Acknowledgments}
The author would like to thank Chiang Cheng-Wei and Chen Jiunn-Wei for reading the manuscript and providing comments.

		\section*{Conflicts of Interest}
{The authors declare no conflict of interest.}  

%
%
%
	
\section*{Use of AI and AI-Assisted Technologies}


{No AI tools were utilized for this paper.}

	\small
	\bibliographystyle{scilight}


\end{document}